\documentclass[twocolumn,aps,prd,preprintnumbers,showpacs,nofootinbibn,notitlepage]{revtex4-1}
\pdfoutput=1     
\usepackage{overpic}
\usepackage{subfigure}
\usepackage[utf8]{inputenc}
\usepackage[english]{babel}
\usepackage{amsmath}
\usepackage{amsfonts}
\usepackage{amssymb}
\usepackage{epsfig}
\usepackage{graphics,psfrag,rotating}
\usepackage{graphicx}
\usepackage{dcolumn}
\usepackage{bm}

\newcommand{\lsim}   {\mathrel{\mathop{\kern 0pt \rlap
  {\raise.2ex\hbox{$<$}}}
  \lower.9ex\hbox{\kern-.190em $\sim$}}}
\newcommand{\gsim}   {\mathrel{\mathop{\kern 0pt \rlap
  {\raise.2ex\hbox{$>$}}}
  \lower.9ex\hbox{\kern-.190em $\sim$}}}

\begin{document}

\title{On the non-attractive character of gravity in $f(R)$ theories}



\author{F. D. Albareti$^{(a)}$}
\author{J.A.R. Cembranos$^{(a)}$}
\author{A. de la Cruz-Dombriz$^{(b,c)}$}
\author{A. Dobado$^{(a)}$}

\affiliation{$^{(a)}$Departamento de F\'{\i}sica
Te\'orica I, Universidad Complutense de Madrid, E-28040 Madrid,
Spain}
\affiliation{$^{(b)}$ Instituto de Ciencias del Espacio (ICE/CSIC) and Institut d'Estudis Espacials de Catalunya (IEEC), Campus UAB, Facultat de Ci\`{e}ncies, Torre C5-Par-2a, 08193 Bellaterra (Barcelona), Spain} 
\affiliation{$^{(c)}$ Astrophysics, Cosmology and Gravity Centre (ACGC) and 
Department of Mathematics and Applied Mathematics, University of Cape Town, Rondebosch 7701, Cape Town, South Africa.} 
%



\date{\today}


%
%


\begin{abstract}

In General Relativity without a cosmological constant
a non-positive contribution from the space-time geometry to Raychaudhuri equation is found provided that particular energy conditions
are assumed and regardless the considered solution of the Einstein's equations. This fact is usually interpreted as a manifestation of the attractive character of gravity. Nevertheless, a positive
contribution to Raychaudhuri equation from space-time geometry should occur since this is the case in an accelerated expanding
Robertson-Walker model for congruences followed by fundamental observers. Modified gravity theories provide the possibility
of a positive contribution although the standard energy conditions are assumed. We address this important issue in the context of $f(R)$ theories,
deriving explicit upper bounds for the contribution of space-time geometry to the Raychaudhuri equation. Then, we examine the
parameter constraints for some paradigmatic $f(R)$ models in order to ensure a positive contribution to this equation. Furthermore, we consider the implications of these upper bounds in the equivalent formulation of $f(R)$ theories as a Brans-Dicke model.

\end{abstract}

\maketitle

\section{Introduction}

Since observational evidence of the accelerated expansion of the Universe was discovered \cite{SIa}, the cosmological evolution as predicted by General Relativity (GR) has been set in doubt. The reason is that a stress-energy tensor possessing strange features needs to be included in the Einstein's
equations in order to account for this cosmic acceleration. This exotic cosmological fluid is usually referred to as dark energy (DE). In its simple
form it is given by a cosmological constant with equation of state $p_{\Lambda}=-\rho_{\Lambda}$.
However, instead of filling the Universe with exotic fluids, a reasonable hypothesis consists in modifying the cosmological field equations assuming
alternative geometrical theories to GR. This approach has received the name of modified gravity theories and has drawn enormous attention in the last
years \cite{Mod_Grav_Theories}.

Some examples are Lovelock theories, whose field equations are second-order differential equations in the metric \cite{Lovelock}; Gauss-bonnet theories inspired in string theory that include a Gauss-Bonnet term in the Lagrangian \cite{GB}; scalar-tensor theories \cite{ST} or vector-tensor theories \cite{VT}, in which gravitational interaction is not only mediated by the standard spin-2 graviton but also by scalar or vector modes respectively; metric theories derived by extra dimensional theories \cite{XD}; supergravity models \cite{sugra}, disformal theories \cite{disformal}, Lorentz violating and CPT breaking models of gravity \cite{LV}; or the so-called $f(R)$ theories, in which our work will be focused.
$f(R)$ theories consist in modifying the Einstein-Hilbert Lagrangian by adding an arbitrary function of the Ricci scalar $R$ (for recent reviews
see \cite{reviewsf(R)}). From this approach, the equations derived from the new action are to be expected as a refinement of the standard Einstein's
equations able to reproduce the correct predictions of GR while explaining the cosmic acceleration. These theories may have strong
effects on small scales, but if some restrictions are imposed, they are able to reproduce the cosmological history while being compatible with local
gravity tests \cite{tests}. It is worth mentioning that the Einstein's equations with cosmological constant $\Lambda$ are a particular case
of these theories with $f(R)=-2\Lambda$.

In fact, the problem with the accelerated expansion of the Universe follows immediately from the consideration of the Friedmann's equations obtained
assuming a Robertson-Walker (RW) cosmological model and a perfect fluid moving along the geodesic congruence followed by the fundamental observers. It
is well-known that stress-energy tensors corresponding to standard fluids cannot be responsible for the accelerated expansion. From a more general point
of view, the Friedmann's equation involving the acceleration of the scale factor results from the Raychaudhuri's equation assuming GR. This equation
provides the expansion rate of a congruence of timelike or null geodesics (see \cite{HE, Wald, Raychaudhuri, Sachs, Ehlers} and recent review \cite{SayanKarandDadhich}). The Raychaudhuri equation plays an important role in the demonstration of the singularities theorems proved by Hawking
and Penrose \cite{HE}. It is usually interpreted that the contribution of space-time geometry to this equation represents the attractive
(or non-attractive) character of gravity. An analysis of this contribution \cite{ACD} showed its geometrical interpretation as the mean curvature
\cite{Eisenhart} in the direction of the congruence. Besides, it can easily be verified that for a RW cosmological model with a negative deceleration parameter
this contribution is positive, i.e.\ the mean curvature in the direction of the fundamental congruence turns out to be positive \cite{ACD}. Hence,
the attractive character of gravity vanishes. From this analysis, it is clear that the accelerated expansion of the Universe may be in conflict with
the attractive character of gravity.

In GR, the attractive character of gravity is assured  by assuming the usual energy conditions \cite{HE, Wald}. Therefore, a positive contribution
to the Raychaudhuri equation from space-time geometry is not attainable in GR provided that these energy conditions hold. Nevertheless, this does not
need to be the case in the context of modified gravity theories. In these theories, even if the usual energy conditions are assumed, a positive
contribution to the Raychaudhuri equation from space-time geometry may be obtained. Moreover, an upper bound to the contribution of space-time geometry
can be provided both in terms of the gravitational model and the metric under consideration. Using this upper bound and assuming the usual energy
conditions, throughout this investigation we shall derive restrictions to $f(R)$ models 
in order to constrain their cosmological viability.

Energy conditions have been widely studied in the literature for different modified gravity theories. The authors of \cite{Santos} generalized energy
conditions for a perfect fluid in $f(R)$ theories by analogy with GR. In \cite{Atazadeh}, the extended energy conditions of $f(R)$ are used to derive
energy conditions in Brans-Dicke theories with a vanishing kinetic term in the Lagrangian using the equivalence between both theories. The energy
conditions have also been studied for $f(R)$ theories with a non-minimal coupling to matter \cite{Bertolami}. In \cite{Garcia}, the same procedure
is applied to Gauss-Bonnet theories. In \cite{Banijamali} and \cite{Zhao}, the authors considered Gauss-Bonnet theories with non-minimal coupling
to matter and derive the corresponding energy conditions. All the aforementioned references followed the formalism first developed in \cite{Santos}.
This generalization of the energy conditions, as the authors of \cite{Santos} themselves first acknowledged remains doubtful since there is no natural
motivation but only an analogy with GR. This extension is in fact only motivated when the new terms appearing in the field equations are identified
with physical fields. Nevertheless, these new terms may be understood as possessing only a geometrical meaning. Thus, there is no reason to assume
any energy conditions on these terms.

Moreover, if these new energy conditions are assumed, the mean curvature in every timelike direction is non-positive by construction and, as already
mentioned, for a RW space-time experiencing an accelerated expansion implies necessarily a positive mean curvature in the direction of the fundamental congruence.
In particular, the stress-energy tensor associated with a cosmological constant $\Lambda$ does not satisfy the usual energy conditions. This fact shows by itself
the limitations of previous investigations assumptions in the most trivial modified gravity Lagrangian beyond pure GR.


This paper is organized as follows: First, in Section \ref{energyconditionsingr}, we summarize the standard energy conditions as considered in GR
and then the role of the Raychaudhuri equation in the theorems of singularities is briefly discussed. Also, the procedure to be developed in the
following section is here sketched. Section \ref{fR_Section} is devoted to introduce the field equations for $f(R)$ theories in the metric formalism as
well as the commonly assumed conditions for $f(R)$ models to be cosmologically viable. Then, in Section \ref{energyconditionsinf(R)}, we assume
the energy conditions in the framework of $f(R)$ theories and present the inequalities that are obtained. We shall proceed by studying
configurations of constant scalar curvature in Section \ref{fR_Examples}. These configurations will enable us to impose some constraints
on the parameters of several relevant $f(R)$ models in order to get a positive contribution to the Raychaudhuri equation. Finally, we conclude our
analysis by presenting our conclusions. In an appendix, we rederived the results obtained in $f(R)$ theories using the alternative representation of these theories as a Brans-Dicke model and we arrive at the same conclusions.

Throughout this study, we use a metric signature ($-,+,+,+$) and our definition of the Riemann tensor is:
\begin{eqnarray}
R_{abc}^{\;\;\;\;\;d}\equiv \partial_b \Gamma^d_{ac}-\partial_a \Gamma^d_{bc}+\Gamma^e_{ac}\Gamma^d_{eb}-\Gamma^e_{bc}\Gamma^d_{ea}\,,
\end{eqnarray}
$R_{ac} \equiv R_{abc}^{\;\;\;\;\;b}$ 
holds for the Ricci tensor and $R=R^{a}_{\ a}$ is the Ricci scalar.
With this convention the usual Einstein's equations yield:
\begin{eqnarray}
R_{ab}-\frac{1}{2}Rg_{ab}\,=\,\frac{8\pi G}{c^4} T_{ab}\,.
\label{Einstein}
\end{eqnarray}
From now on, we shall adopt  $c=G=1$. Furthermore, the stress-energy tensor corresponding to a perfect fluid is:
\begin{eqnarray}
T_{ab}=\rho\, \xi_a \xi_b +p \left(g_{ab}+\xi_a \xi_b\right)\,.
\label{fluido}
\end{eqnarray}

\section{Energy conditions in General Relativity}
\label{energyconditionsingr}

The Raychaudhuri equation for timelike geodesics can be expressed as \cite{Wald,Raychaudhuri}
\begin{eqnarray}
\frac{\text{d}\theta}{\text{d}\tau}=-\frac{1}{3}\,\theta^2-\sigma_{ab}\sigma^{ab}+\omega_{ab}\omega^{ab}-R_{ab}\xi^a\xi^b\,,
\label{ray}
\end{eqnarray}
where $\theta$, $\sigma_{ab}$ and $\omega_{ab}$ are respectively the expansion, shear and twist of the congruence of timelike geodesics generated by the tangent vector field $\xi^a$ and $\tau$ is an affine parameter. On the other hand, the analogous equation for null geodesics becomes \cite{Wald,Sachs}
\begin{eqnarray}
\frac{\text{d}\hat{\theta}}{\text{d}\lambda}=-\frac{1}{2}\,\hat{\theta}^2 -\hat{\sigma}_{ab}\hat{\sigma}^{ab}+\hat{\omega}_{ab}\hat{\omega}^{ab}-R_{ab} k^a k^b\,,
\label{raynull}
\end{eqnarray}
where $k^a$ is the tangent vector field to a congruence of null geodesics and $\lambda$ is an affine parameter. Let us recall that \eqref{ray} and \eqref{raynull} are geometrical identities, thus they hold independently of the gravitational theory assumed.

In this investigation we are interested in the contribution of space-time geometry to the previous equations, i.e., $-R_{ab}\xi^a\xi^b$ and
$-R_{ab} k^a k^b$. If a particular form for the metric tensor is assumed a priori, the Ricci tensor can be determined and then those contributions
can be directly studied as it is the case for a RW metric \cite{ACD} without considering any underlying gravitational theory. However,
the metric tensor is in general unknown from the beginning, thus an expression for the Ricci tensor is not at our disposal. In the latter scenario, the problem
can be nonetheless tackled by using the field equations to obtain information about $-R_{ab}\xi^a\xi^b$ and $-R_{ab} k^a k^b$. This procedure leads
to considering conditions to be imposed on the stress-energy tensor $T_{ab}$, the so-called energy conditions.
Let us thus revise these conditions and some of their implications in GR both for timelike and null vectors.


\subsection{Timelike vectors}


The energy density of matter as measured by an observer with velocity $\xi^a$, is $T_{ab}\xi^a\xi^b$. It is reasonable that this density would be non-negative. This requirement is known as the weak energy condition (WEC)
\begin{eqnarray}
T_{ab}\xi^a\xi^b\geq 0\ \ \ \ \ \ \ \ \ \ \ \ \ \ \ \ \ \ \ \text{WEC}\,.
\label{WEC}
\end{eqnarray}
Moreover, the dominant energy condition (DEC) ensures that the speed of the flux of energy is less than the speed of light, yielding
\begin{eqnarray}
T_{ab}\xi^a\, T^{bc}\xi_c \leq 0\ \ \ \ \ \ \ \ \ \ \ \ \ \ \ \text{DEC}\,,
\label{DEC}
\end{eqnarray}
which expresses that the flux of energy, i.e.\ $-T^{ab}\xi_a$, is a timelike vector where the minus sign appears because we have chosen signature $(-,+,+,+)$.
Furthermore, we are mainly interested in 
the expression $R_{ab}\xi^a\xi^b$. Using the usual Einstein's equations \eqref{Einstein} and the subsequent relation between the Ricci scalar and the trace of the stress-energy tensor, i.e.\ $R=-8\pi T$, we obtain
\begin{eqnarray}
\begin{aligned}
R_{ab}\xi^a\xi^b\, &=8\pi\left(T_{ab}-\frac{1}{2}Tg_{ab}\right)\xi^a\xi^b \\ &=
8\pi\left(T_{ab}\xi^a\xi^b+\frac{1}{2}T\right)\,.
\end{aligned}
\label{Rab}
\end{eqnarray}
It is customary to assume the positive sign of the r.h.s. of this equation since 
a distribution of standard matter would not result in a stress-energy tensor with pressure so large and negative as to make this member negative. This statement can be understood after replacing expression  \eqref{fluido}  
in \eqref{Rab}. Hence,  stress-energy tensors for standard matter fluids 
satisfy the so-called strong energy condition (SEC)
\begin{eqnarray}
T_{ab}\xi^a\xi^b\geq-\frac{1}{2}T,\ \ \ \ \ \ \ \ \ \ \ \text{SEC}\,.
\label{SEC}
\end{eqnarray}
It is known that both dust matter and radiation satisfy the SEC.
For a discussion about cases where this condition does not hold see \cite{HE}. In particular, a stress-energy tensor corresponding to a cosmological constant $\Lambda$ fluid does not fulfill the SEC. We will discuss this case at the end of the section to avoid losing continuity in the discussion.
Therefore, the SEC requires
\begin{eqnarray}
R_{ab}\xi^a\xi^b\geq 0\,,
\label{R0}
\end{eqnarray}
which may be interpreted, because of asserting a non-positive contribution to Raychaudhuri equation, as a manifestation of the attractive character of gravity. It follows that the mean curvature \cite{ACD, Eisenhart} in every timelike direction defined by
\begin{eqnarray}
{\cal M}_{\xi^a}\,\equiv\,-R_{ab}\xi^a\xi^b
\label{Mean_curvature}
\end{eqnarray}
is negative or zero in GR provided that the SEC is assumed.

The usefulness of the Raychaudhuri equation
in the singularity theorems is based upon the following result: if one chooses a
congruence of timelike geodesics  whose tangent vector field is locally hypersurface-orthogonal, then $\omega_{ab}=0$ for all the congruence (as a consequence of Frobenius' theorem \cite{Wald}) is obtained.
The term $\sigma_{ab}\sigma^{ab}$ is non-negative and whenever $R_{ab}\xi^a\xi^b \geq 0$ is assumed, then
\begin{eqnarray}
\frac{\text{d}\theta}{\text{d}\tau}+\frac{1}{3}\theta^{2}\leq 0\,,
\end{eqnarray}
which implies
\begin{eqnarray}
\theta^{-1}(\tau) \geq \theta^{-1}_0 +\frac{1}{3}\tau\,.
\end{eqnarray}
This inequality 
tells us that a congruence initially converging ($\theta_0\leq 0$) will converge until zero size in a finite time $\tau \leq 3/\lvert \theta_{0} \rvert$, or in a reversed sense, if the congruence is initially diverging $\theta_{0} \geq 0$ it was focused until zero size in the past. This result is important in proving the theorems concerning singularities of Hawking and Penrose \cite{HE, Wald}. Very often it is claimed that these theorems require energy conditions to hold, since for instance the SEC as we have just seen implies $R_{ab}\xi^a\xi^b\geq0$. However, these theorems are essentially mathematical theorems independent of the gravitational theory. Energy conditions are necessary for these theorems to hold if and only if GR is assumed. Otherwise, the requirement would be $R_{ab}\xi^a\xi^b \geq 0$ for every non-spacelike vector. In Section \ref{energyconditionsinf(R)}, we shall assume the usual energy conditions in the framework of $f(R)$ theories but the sign of $R_{ab}\xi^a\xi^b$ will remain in principle undetermined.


\subsection{Null vectors}


Let us now consider a congruence of null geodesics. Just by replacing $\xi^{a}\rightarrow k^{a}$ in equation \eqref{Rab}, one gets
\begin{eqnarray}
R_{ab}k^a k^b=8\pi\left(T_{ab}-\frac{1}{2}Tg_{ab}\right)k^a k^b=8\pi T_{ab} k^a k^b.
\label{Rabnull}
\end{eqnarray}
Hence, the so-called null energy condition (NEC)
\begin{eqnarray}
T_{ab} k^a k^b\geq 0\ \ \ \ \ \ \ \ \ \ \ \ \ \ \ \ \ \ \ \text{NEC}\,,
\label{NEC}
\end{eqnarray}
implies that $R_{ab} k^a k^b$ will be non-negative. NEC is fulfilled by continuity if the SEC is assumed, but it is also fulfilled by imposing the WEC. Then, the assumptions for a congruence of null geodesics to focus are weaker than those for a congruence of timelike geodesics sketched above. Reasoning in the same way as before,
if a congruence of null geodesics is initially converging $\hat{\theta}_{0} < 0$ it will converge until zero size in a finite time $\tau \leq 2/\lvert \hat{\theta}_{0} \rvert$ or in the reversed sense.

The convergence of timelike and null geodesics in a finite time -- in the future or in the past -- is ensured in GR under the assumptions of the SEC and the NEC respectively. This result is usually known as \textit{the geodesic focusing theorem} \cite{HE, Wald}.



\subsection{Beyond General Relativity}

If one considers a congruence whose tangent vector field is locally hypersurface-orthogonal, this means that, on the one hand $\omega_{ab}=0$ and, on the other hand, the r.h.s. of the Raychaudhuri equation for timelike geodesics \eqref{ray}, has only non-positive contributions of the parameters of the congruence ($- \theta^2/3$ and $-\sigma_{ab} \sigma^{ab}$).
%
If the observed acceleration of the Universe needs to be explained, a positive contribution of the space-time given by ${\cal M}_{\xi^a}>0$ 
is required at least for some directions $\xi^a$. 
This is unfeasible in GR if the SEC is satisfied as can be seen from \eqref{R0}. This opens two ways
of circumventing the unavoidable attractive character of gravity in GR: either
%
to suppose that the SEC is not satisfied
or to modify the Einstein's equations.
%
An analogous discussion may be done for null geodesics replacing the SEC by the NEC.

In this work we shall consider the second scenario: the SEC will be located in a privileged place with respect to the Einstein's equations. We are thus assuming that standard matter satisfies the SEC and that the possibility of a positive contribution to Raychaudhuri equation through the mean curvature ${\cal M}_{\xi^a}$ 
must be obtained from the $f(R)$ modified field equations. By proceeding in this way, an inequality involving $R_{ab} \xi^a \xi^b$ and
terms depending on the gravitational theory under consideration will provide us an upper bound to the contribution of space-time geometry ${\cal M}_{\xi^a}$. This bound will allow
us to derive some restrictions on the $f(R)$ models in order to get ${\cal M}_{\xi^a}$ positive, or equivalently $R_{ab} \xi^a \xi^b$ negative.

As it was mentioned above,   the cosmological constant stress-energy tensor does not satisfy the SEC and
should not be regarded as an stress-energy term but as a particular $f(R)=-2\Lambda$ model.
\section{$f(R)$ theories}
\label{fR_Section}

%
%
%
%

Let us consider the total action
\begin{eqnarray}
S=S_{grav}+S_{matter}\,,
\label{S_total}
\end{eqnarray}
i.e.\ a gravitational action plus a matter action term that includes all matter fields. The modification of $f(R)$ theories to GR consists in assuming $S_{grav}$ of the form
\begin{eqnarray}
S_{grav}=\frac{1}{16 \pi }\int \left(R+f(R)\right)\,\sqrt{\mid g\mid}\text{d}^{4}x\,,
\label{action}
\end{eqnarray}
where $f(R)$ is an arbitrary function of $R$ and $g$ is the determinant of the metric. 
%
%
The variation of \eqref{S_total} with respect to the metric tensor 
yields 
\begin{eqnarray}
 R_{ab}(1+f'(R)) - \frac{1}{2}\,g_{ab}\,\left(R+f(R)\right) \nonumber \\-\left(\nabla_a \nabla_b-g_{ab}\Box\right)f'(R)&=&8\pi \,T_{ab}\,.
\label{ec_campo}
\end{eqnarray}
where the stress-energy tensor is defined as
\begin{eqnarray}
T_{ab} \equiv -\frac{2}{\sqrt{\mid g\mid}}\frac{\delta S_{matter}}{\delta
g^{ab}}\,.
\label{tensor_fR}
\end{eqnarray}
Equations (\ref{ec_campo}) are obtained in the metric formalism, i.e.\ the connection is assumed to be Levi-Civita connection.
%
$f(R)$ theories have been proved to be able to reproduce
the cosmological history from inflation to the current accelerated expansion era. For instance, it has been showed that the evolution of the Universe can be reproduced with certain $f(R)$ functions \cite{Dombriz-Dunsby-Odintsov}.

Let us remark here that equations \eqref{ec_campo} are fourth order differential equations. This is the reason of some strong instabilities that arise for certain $f(R)$ models, such as the Dolgov-Kawasaki instability for the model $f(R)=-\mu^4/R$ \cite{Dolgov-Kawasaki}. Moreover, there is a general instability known as Ostrograski instability associated with Lagrangians that contain non-linear second derivatives terms. However, it has been proved that one can avoid Ostrograski instabilities in $f(R)$ theories ({\it cf.} \cite{Clifton}). The Cauchy problem in $f(R)$ theories has also been considered \cite{Cauchy} where the authors concluded that the problem is well-posed in the metric formalism.
Further details as well as local and cosmological tests for $f(R)$ theories can be seen in  \cite{reviewsf(R), tests} among others.


\subsection{Viability conditions of $f(R)$ theories}

Some constraints are usually imposed on the $f(R)$ functions in order to provide consistent theories of gravity. Three
of those conditions that shall be taken into account in our study are:

\begin{enumerate}

\item $1+f'(R)>0$. This condition is imposed in order to ensure a positive effective gravitational constant $G_{eff} \equiv G/\left(1+f'(R)\right)$. It
means that the main part of the contribution to the Einstein's equations conserves the sign \cite{Pogosian}. This condition also guarantees the non-tachyonic character of the standard graviton.

\item $f''(R)\geq 0$. It ensures a stable gravitational stage. It is directly related to the presence of a positive mass in a high curvature regime for
the scalar mode associated with this type of theories \cite{fRcha}.

\item $f(R)/R \rightarrow 0$ as $R \rightarrow \infty$. This last condition ensures to get GR behavior at early times. This way, we recover the correct predictions of GR about Big Bang nuclesynthesis and the CMB. However, if we are only interested in analyzing models for cosmic acceleration, this last condition is not required.

\end{enumerate}

In the following discussion, the first two conditions will be assumed. The last condition will also be discussed for completeness.

At this stage, let us remember that there also exist several constraints
for the value of $|f'(R_0)|$ where $R_0$ holds either for
the current or past background curvature ({\it cf}. \cite{Referee}). 
However, these constraints must be carefully interpreted since they are obtained
under several assumptions, depend on the astrophysical curvature under consideration
and in general, they are model-dependent. A particular $f(R)$ theory for a particular
value of parameters needs a proper analysis that is beyond the scope of this work.
In any case, although these constraints are not directly applicable in our line of study,
it is interesting to keep in mind that they may be important,
and we will show for reference upper bounds from cosmological
tests that can be found in the present literature \cite{Referee}. Namely,
\begin{eqnarray}
|f'(R_0)| < 0.35 \;\;;\;  |f'(R_0)| < 0.07
\label{constraints_fR}
\end{eqnarray}
according to integrated Sachs-Wolfe effect measured by CMB temperature spectrum and correlations thereof with foreground galaxies respectively \cite{Referee}.
Finally, in accordance with the authors in \cite{Hu}, let us
stress that it is worthwhile to consider
$f(R)$ theories as effective cosmological theories valid for
an adequate range of curvatures regardless the small-scale
tests of gravity. Thus, they may provide an adequate
phenomenological framework to describe new kind of phenomena
which deviate from GR behavior at large enough scales.

\section{Energy conditions in $f(R)$ theories}
\label{energyconditionsinf(R)}

In this section we are interested in analogous equations to \eqref{Rab} and \eqref{Rabnull} when extended to $f(R)$ theories. Thus, by imposing the usual energy conditions, inequalities involving the terms $R_{ab}\xi^a\xi^b$ (or $R_{ab} k^a k^b$) and $f(R)$ extra geometrical terms
will be obtained. These inequalities will provide us an upper bound for the contribution of space-time geometry to Raychaudhuri equation for timelike geodesics \eqref{ray} and for null geodesics \eqref{raynull}.

The usual approach in literature \cite{Santos,Atazadeh,Bertolami,Garcia,Banijamali,Zhao} 
consisted of defining an effective stress-energy tensor by analogy with GR that includes the new geometrical terms in order to obtain an expression for $R_{ab}\xi^a\xi^b$. Then, generalized energy conditions on this effective tensor were imposed. By this procedure, a negative or zero contribution to Raychaudhuri equation from the term ${\cal M}_{\xi^a}=-R_{ab}\xi^a\xi^b$  for every timelike direction $\xi^a$ is
obtained whenever these analogous energy conditions hold. Nevertheless, as we have already mentioned,
${\cal M}_{\xi^a}>0$ is satisfied for almost all timelike directions in a RW cosmological model with the present value of the deceleration parameter $q_{0}$ \cite{ACD}. Therefore, if those extended energy conditions hold, the present accelerated expansion of the Universe cannot be accommodated.

%

\subsection{Inequalities derivation}
\label{Introduction_energyconditionsinf(R)}

Let us first take the trace of equation \eqref{ec_campo} that can be recast as
\begin{eqnarray}
-8\pi T=R\,(1-f'(R))+2f(R)-3\Box f'(R)\,.
\label{R}
\end{eqnarray}
%
Therefore, from \eqref{ec_campo} together with \eqref{R} one gets
\begin{eqnarray}
&&R_{ab}\left(1+f'(R)\right)-\frac{1}{2}g_{ab}\left(R f'(R)-f(R)+3\Box f'(R)\right)\nonumber\\
&&-\left(\nabla_a\nabla_b-g_{ab}\Box\right)f'(R)
\,=\,8\pi \left(T_{ab}-\frac{1}{2} T g_{ab}\right).
\label{ec_campoGR}
\end{eqnarray}
Contracting the last equation with $\xi^a\xi^b$, where $\xi^a$ is a normalized timelike vector, $\xi^a \xi_a=-1$, we get
\begin{eqnarray}
&&R_{ab}\xi^a\xi^b\left(1+f'(R)\right)-\left(\xi^a\xi^b\nabla_a\nabla_b-\frac{1}{2}\Box\right)f'(R) \nonumber \\
&&+\frac{1}{2}\left(Rf'(R)-f(R)\right)\,=\,8\pi \left(T_{ab}\xi^a\xi^b+\frac{1}{2} T \right)\,.
\label{contractedtimelike}
\end{eqnarray}
If we consider a null vector $k^a$ instead of a timelike vector, then multiplying \eqref{ec_campoGR} by $k^a k^b$ the result becomes
\begin{eqnarray}
&&R_{ab}k^a k^b\left(1+f'(R)\right)- k^a k^b\nabla_a\nabla_b f'(R) \,=\,8\pi T_{ab} k^a k^b\,.\nonumber\\
&&
\label{contractednull}
\end{eqnarray}
At this stage, let us impose the SEC and the NEC to the standard cosmological fluids in the expressions \eqref{contractedtimelike} and \eqref{contractednull} respectively. After some manipulations, they become
%
%
%
%
%
\begin{eqnarray}
&&R_{ab}\xi^a\xi^b \geq 
\frac{1}{2(1+f'(R))} \nonumber\\
&&\times \left[f(R)-R f'(R)+\left(2\xi^a\xi^b\nabla_a\nabla_b-\Box\right)f'(R)\right]
\label{I}
\end{eqnarray}
and
\begin{eqnarray}
R_{ab}k^a k^b\geq 
\frac{1}{1+f'(R)} k^a k^b\nabla_a\nabla_b f'(R)\,,
\label{II}
\end{eqnarray}
where in both expressions $1+f'(R)$ was assumed to be positive in order
to guarantee $G_{eff}\equiv G/(1+f'(R))>0$. Let us remind that our conclusions will be based upon this requirement.
%
As a result of \eqref{I} and \eqref{II}, one can conclude that although the SEC (NEC) have been assumed, in $f(R)$ theories the sign of
$R_{ab}\xi^a\xi^b$ ($R_{ab}k^a k^b$)
cannot be determined a priori.
Thus, if a certain model $f(R)$ renders negative the right-hand side (r.h.s.) of \eqref{I} or \eqref{II}, some freedom remains for $R_{ab}\xi^a \xi^b$ or $R_{ab} k^a k^b $ to be negative which may be interpreted as a repulsive force.
%
Reminding now the definition \eqref{Mean_curvature},  the inequalities \eqref{I} and \eqref{II} can be cast in the following way
\begin{eqnarray}
&&{\cal M}_{\xi^a}=-R_{ab}\xi^a\xi^b \leq \frac{-1}{2(1+f'(R))}\nonumber \\
&& \times \left[ f(R)-R f'(R) + \left(2\xi^a\xi^b\nabla_a\nabla_b - \Box\right)f'(R)\right],
\label{Iupper}
\end{eqnarray}
\begin{eqnarray}
-R_{ab}k^a k^b \leq \frac{-1}{1+f'(R)} k^a k^b\nabla_a\nabla_b f'(R)\,;
\label{IIupper}
\end{eqnarray}
that provide upper bounds to the contribution of space-time geometry to the Raychaudhuri equation for timelike  and null geodesics respectively.

Let us stress that in vacuum the inequalities \eqref{I} and \eqref{II} -- or equivalently
\eqref{Iupper} and \eqref{IIupper} -- get saturated since the SEC/NEC energy conditions are trivially
saturated. Consequently, if a $f(R)$ model renders the r.h.s.\ of \eqref{Iupper} and \eqref{IIupper} positive in vacuum scenario, then a positive contribution to the Raychaudhuri equation is automatically obtained for timelike geodesics and null geodesics respectively.

Let us focus on a short example. The Einstein's equations with a cosmological constant $\Lambda$ become
\begin{eqnarray}
R_{ab}-\frac{1}{2}Rg_{ab}+\Lambda g_{ab}=8\pi T_{ab}\,.
\end{eqnarray}
As we have previously noted, GR with a cosmological constant is equivalent to take $f(R)=-2\Lambda$. In this trivial case, the
inequalitites \eqref{Iupper} and \eqref{IIupper} become
\begin{eqnarray}
{\cal M}_{\xi^a} \leq \Lambda \ \ \ ,\ \ \ \ \
-R_{ab}k^a k^b \leq 0\,.
\label{shortexample}
\end{eqnarray}
The first inequality tells us that in the case of timelike geodesics a positive contribution to the Raychuadhuri equation from space-time geometry ${\cal M}_{\xi^a} > 0$ is possible provided that $\Lambda > 0$, which corresponds to the correct sign of $\Lambda$ to provide cosmic acceleration.

Hence, it may be thought that a simple criterion to decide when a $f(R)$ model is able to render accelerated expansion of the Universe has been obtained.
Nonetheless,  expressions \eqref{I} and \eqref{II} need to be evaluated at the solutions for \eqref{ec_campo}  so the problem remains cumbersome.
However, 
such a problem is absent when space-time configurations
with constant scalar curvature $R_{0}$ are considered.

%
%



\subsection{Constant scalar curvature solutions}
\label{constantsolutions}

%
%

Space-times both in vacuum and GR cosmological scenarios
when studied at late times, with both radiation and dust being negligible with regard to a cosmological constant,
are maximally symmetric, i.e.\ they possess a constant Gaussian curvature $K_{0}$. This implies a constant scalar curvature $R_{0}=12K_{0}$ (but the reverse is not generally true).
For this reason, it may be expected that solutions of constant scalar curvatures will be recovered at late times by physically viable $f(R)$ models.


Since the covariant derivatives of $f'(R)$ in solutions of constant scalar curvature ($R=R_0$) are zero, expressions \eqref{I} and \eqref{II} result respectively in
\begin{eqnarray}
R_{ab}\xi^a\xi^b \geq \frac{f(R_0)-R_{0}f'(R_{0})}{2(1+f'(R_0))}
\label{Ib}
\end{eqnarray}
and
\begin{eqnarray}
R_{ab}k^a k^b \geq 0\,.
\label{IIb}
\end{eqnarray}
A remarkable result follows from the inequality \eqref{IIb}: In $f(R)$ theories, the condition for the null geodesic focusing theorem to hold, namely $R_{ab}k^ak^b \geq 0$, is satisfied in space-times of constant scalar curvature provided that the NEC is assumed as given by \eqref{NEC}. It is worth noticing that this result does not depend upon the sign of $R_{0}$ nor upon the $f(R)$ model under consideration. Moreover, the NEC is only assumed in the standard stress-energy tensor for matter, not in the {\it effective} one usually defined after gathering all the new terms of the modified Einstein equations.
Since  the holographic principle \cite{Bousso} makes use of the null geodesic focusing theorem in order to ensure that light-sheets will eventually end, the previous result
is of extraordinary importance when studying this principle in $f(R)$ theories.

In the rest of this investigation we shall focus on timelike geodesics. Therefore, the r.h.s.\ of \eqref{Ib} must be negative in order to allow $R_{ab}\xi^a\xi^b < 0$ or equivalently ${\cal M}_{\xi^a} > 0$. 
Thus, ${\cal M}_{\xi^a}$ be bounded from above. 
Hence we impose
\begin{eqnarray}
\frac{f(R_0)-R_{0}f'(R_{0})}{2(1+f'(R_0))} < 0\,,
\end{eqnarray}
and provided that $G_{eff}>0$, we get
\begin{eqnarray}
f(R_0)-R_{0}f'(R_{0}) < 0\,.
\label{lacondicion}
\end{eqnarray}
%
If we now consider the equation \eqref{R} in vacuum ($T=0$) for constant scalar curvature solutions, 
the value of $R_0$ satisfies
\begin{eqnarray}
R_0=\frac{-2f(R_0)}{1-f'(R_0)}\,,
\label{ec_curvatura_constante}
\end{eqnarray}
which is an algebraic equation relating $R_{0}$ with the parameters of the $f(R)$ model under study. Although in general this equation cannot be solved analytically,
there exist some $f(R)$ models for which a closed solution depending upon the parameters of the model can be found. Using the equation \eqref{ec_curvatura_constante} in \eqref{lacondicion} one gets
\begin{eqnarray}
\frac{f(R_0)}{1-f'(R_0)} < 0\,,
\label{vacio}
\end{eqnarray}
that together with \eqref{ec_curvatura_constante}, implies
\begin{eqnarray}
R_0 > 0\,.
\label{Rmayor0}
\end{eqnarray}
%
In the pathological case $1-f'(R)=0$, the equation \eqref{R} reads
\begin{eqnarray}
f(R_{0})=0\,.
\end{eqnarray}
then, \eqref{lacondicion} results in
\begin{eqnarray}
R_{0}\,f'(R_{0})>0\,,
\end{eqnarray}
which is equivalent to $R_{0}>0$ since in this case
$f'(R_{0})=1$.
Hence, a positive contribution to the Raychaudhuri equation from the space-time geometry ${\cal M}_{\xi^a}$ 
for every timelike direction is obtained provided that $R_{0}>0$. This condition will 
constrain the parameters of different $f(R)$ models as will be seen in the next section.

In fact, there exists another straightforward way of getting \eqref{vacio} as follows: The solutions of \eqref{ec_campo} in vacuum with constant scalar curvature imply
\begin{eqnarray}
R_{ab}\,\left(1+f'(R_0)\right)-\frac{1}{2}\,g_{ab}\,\left(R_0+f(R_0)\right)=0\,
\end{eqnarray}
and consequently
\begin{eqnarray}
R_{ab}=\frac{1}{2}\frac{R_{0}+f(R_0)}{1+f'(R_0)}\,g_{ab}=\frac{R_0}{4}g_{ab}\,,
\label{Riccialfa}
\end{eqnarray}
where \eqref{ec_curvatura_constante} has been used in the last equality. It means that the allowed $f(R)$ space-times with
constant scalar curvature in vacuum 
are Einstein spaces \cite{Eisenhart}.
%
Thus, if a negative value of $R_{ab}\xi^a\xi^b$ is required in order to have ${\cal M}_{\xi^a}>0$ 
the condition $R_0>0$
\ needs to be accomplished.
However, one must keep in mind that the inequality \eqref{lacondicion} is more general since it allows us to have ${\cal M}_{\xi^a} > 0$ even
for non-vacuum scenarios.

 If one considers a maximally symmetric space-time, the condition \eqref{Rmayor0} implies that in order to guarantee a
 positive contribution to the Raychadhuri equation the space-time must be de Sitter ($R_0>0$).
As it is widely known, a de Sitter space-time may be foliated by 
constant curvature spacelike hypersurfaces which may be of positive, negative or zero Gaussian curvature. 
Therefore, a de Sitter space-time provides us with a common language in order to describe accelerated expanding Universe with close, open o flat spacelike hypersurfaces.
The parameters constraints for different $f(R)$ models explored in the following guarantee
that such a space-time exists as a solution of the modified Einstein's equations \eqref{ec_campo}.

However, if one is interested in constant scalar curvature space-times which are not maximally symmetric the following differential equation
\begin{eqnarray}
\frac{\ddot{a}}{a}+\left(\frac{\dot{a}}{a}\right)^2 +\frac{k}{a^2}=\frac{R_{0}}{6}\,,
\end{eqnarray}
where dot stands for derivatives with respect to cosmic time; needs to be solved to get a general expression for the evolution of the scale factor of a RW metric in these space-times. In the case of flat spacelike sections, after a change of variables this equation results in a Riccati-type ODE that may be solved using standard methods. Hence, a constant scalar curvature space-time may give rise to different dynamical evolution of an expanding Universe.

\section{$f(R)$ models}
\label{fR_Examples}

Let us now study some $f(R)$ models in vacuum to illustrate the previous results:
%
\begin{itemize}
	
	\item {\bf Model I} $f(R)=\alpha\left|R\right|^\beta$
	
	\smallskip

\ This model encompasses a wide variety of proposals available in the literature.
The case $\beta<1$ and in particular $\beta=-1$ 
was proposed in \cite{Carroll} as a possible mechanism to provide cosmological acceleration, although it is currently
excluded due to the Dolgov-Kawasaki instability \cite{Dolgov-Kawasaki} extended in \cite{485} 
for functions $f(R)$ modifying gravity in the infra-red limit. Extension \cite{485} proved that  for negative exponents $\beta$, models with $\alpha<0$
are not stable. 
On the other hand, the $\beta=2$ case has been proposed both as a viable inflation
candidate by Starobinsky \cite{Starobinsky} and as a dark matter model \cite{Cembranos}.  In this last
reference, the $\alpha$ parameter definition reads
$\alpha = (6 m_2^0)^{-1}$ and the minimum value allowed for $m_0$ is computed as
$m_0 = 2.7 \times 10^{-12}$ GeV at $95\%$ confidence level, i.e.  $\alpha\leq 2.3 \times 10^{22} {\rm GeV}^{-2}$ \cite{26}.

\ Concerning other exponents, the literature is extensive \cite{varia_Rn} dealing in general
with the value of the effective mass in the scalar degree of freedom of this class of models, which is thought to be either
too small for validity of gravitational physics in the solar system, or imaginary, leading to some of the
instabilities alluded by the violation of condition  $f''(R)\geq0$.

%

\ For this model, the curvature scalar can be obtained from \eqref{ec_curvatura_constante} yielding
\begin{eqnarray}
R_{0}=\pm \left[\frac{\pm1}{\alpha (\beta -2)}\right]^{\frac{1}{\beta-1}}\,,
\end{eqnarray}
where the sign depends upon the sign of $R_{0}$ because of the derivative of the absolute value (plus signs for $R_{0}>0$ and minus signs for $R_{0}<0$). There is also a trivial solution with $R_{0}=0$ which is of no interest for our discussion. Since we are interested in $R_{0}>0$ we take the expression with the plus signs. In order to a positive constant scalar curvature exist the parameters must obey
%
%
%
\begin{eqnarray}
\alpha(\beta-2)>0.
\end{eqnarray}
Note that for $\beta=2$ the only constant scalar curvature in vacuum is $R_0=0$.

\begin{figure}
	\begin{center}
	\resizebox{8cm}{8cm}{
		\includegraphics[bb=0 0 350 350]{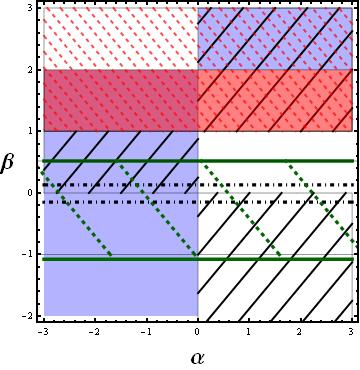}}
	\caption{
	\label{fig:primermodelovacio}
	\footnotesize
	$(\alpha,\,\beta)$ plane for Model I 
	in vacuum: The pairs $(\alpha,\beta)$ lying in the blue region are those that satisfy $\alpha(\beta-2)>0$ and consequently $R_0 > 0$. The black meshed zone fulfills $f''(R_{0}) \geq 0$ which is a stability condition commented in Section \ref{fR_Section}. The red zone does not satisfy the condition $G_{eff}=G/(1+f'(R_{0}))>0$ and therefore the inequality \eqref{vacio} is not valid there as commented in the previous section. Finally, the red meshed zone does not satisfy $f(R)/R \rightarrow 0$ as $R \rightarrow \infty$, thus for these parameters one cannot recover GR behavior at early times. The parameters that provide a positive contribution from the space-time geometry, i.e.\ ${\cal M}_{\xi^a}=-R_{ab}\xi^a\xi^b > 0$, to the Raychaudhuri equation for congruences of timelike geodesics are those of the blue zone excluding the red zone about which no statement can be done with our discussion. We have also added current cosmological constraints on the value of $|f'(R_{0})|$, the green meshed region is compatible with $|f'(R_{0})|<0.35$ and the region between the dashed black lines is compatible with $|f'(R_{0})|<0.07$.}
	\end{center}
\end{figure}

\ Regions where 
the conditions  $\alpha(\beta-2)>0$ and $f''(R_{0}) \geq 0$
hold are plotted in Figure \ref{fig:primermodelovacio}.
The region where $G_{eff}=G/(1+f'(R_{0}))>0$ does not hold is also represented.
Since this viability condition of $f(R)$ has been assumed in deriving the inequality \eqref{vacio}, our discussion is not valid for the values of the parameters falling in that region.
Let us stress that the conditions $f''(R_{0}) \geq 0$ and $G_{eff} > 0$ are evaluated in the corresponding value of the constant scalar curvature $R_{0}$ which depends on the parameters.
It means that these conditions will be satisfied by the parameters that fall in the corresponding regions in the case of constant scalar curvature solutions, not for every solution of the equation \eqref{ec_campo}. This consideration remains valid for all the models studied in this section.

\ In Figure \ref{fig:primermodelovacio} one can also see that for $\beta > 0$ there are regions where both requirements, namely $R_{0} > 0$ and $f''(R_{0}) \geq 0$, hold. For $\alpha < 0$, $\beta$ must be restricted to the interval $(0,1)$; on the contrary, for $\alpha > 0$, it must be $\beta > 2$; in order both requirements to hold.

\ Finally, if one also considers the third condition of viability of Section \ref{fR_Section}, the values of the parameters are highly constrained. However, for models with $\alpha < 0$ and $0 < \beta < 1$, it is still possible to get a constant scalar curvature solution which provides a positive contribution to Raychaudhuri equation while the model being stable and reproducing GR behavior at early times. Moreover, some of these region are compatible with currents cosmological constraints on the value of $|f'(R)|$ discussed at the end of Section \ref{fR_Section}.

\bigskip

  \item {\bf Model II} $f(R)=R^\alpha \exp(\beta/R)-R$

\begin{figure}
	\begin{center}
	\resizebox{8cm}{8cm}{
		\includegraphics[bb=0 0 350 350]{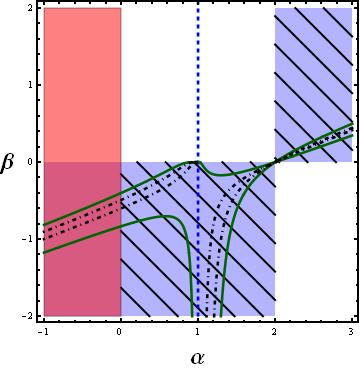}}
	\caption {\label{fig:segundomodelovacio}\footnotesize
$(\alpha,\,\beta)$ plane for Model II 
	in vacuum: For the parameters in the blue zone a positive constant scalar curvature exists. The condition $f''(R_{0})\geq0$ is fulfilled in the meshed zone. Let us remember that for this model the condition $f''(R_{0})\geq0$ depends on the sign of $R_{0}$ and the value of $\alpha$. For simplicity, the condition $f''(R_{0})\geq0$ is plotted after assuming $R_{0}>0$ since this is the case in which we are interested. Moreover, for $R_{0}>0$ the condition $f''(R_{0})\geq0$ does not depend on the value of $\alpha$. In the red region the condition $G_{eff}=G/(1+f'(R_{0}))>0$ does not hold and consequently the inequality \eqref{vacio} does not apply there. For $\alpha=1$, represented by the dashed blue line, the model recovers GR behavior at early times. Finally, the same upper bounds on the value of $|f'(R_{0})|$ considered in the previous model, namely $|f'(R_{0})|<0.35$ and $|f'(R_{0})|<0.07$, are plotted. These restrictions are satisfied between the green lines and dashed black lines respectively. Let us recall that we have only plotted these contraints on the region where $R_{0}>0$ for simplicity.}
	\end{center}
\end{figure}

  \smallskip

\ This model was discussed for $\alpha=1$ in \cite{Amendola} and more recently in \cite{Black holes}. 
By proceeding as for the previous model, one gets from equation \eqref{ec_curvatura_constante}
\begin{eqnarray}
R_0=\frac{\beta}{\alpha -2}\,.
\end{eqnarray}
Thus, in order to ensure a positive contribution to the Raychaudhuri equation from the space-time geometry, i.e.\ ${\cal M}_{\xi^a}=-R_{ab}\xi^a\xi^b > 0$ one must impose
\begin{eqnarray}
\frac{\beta}{\alpha-2} > 0\,.
\end{eqnarray}

Note that for $\alpha=2$, the only constant scalar curvature vacuum solution is $R_0=0$. The same conditions as for the previous model are plotted in Figure \ref{fig:segundomodelovacio}. Let us remind that the conditions $G_{eff}>0$ and $f''(R_{0}) \geq 0$ are evaluated in $R_{0}$ which depends on the parameters of the model.

\ Let us stress 
that for this case the value of $f''(R_{0})$ depends on the sign of $R_{0}$. Since we are interested in the region where $R_{0} > 0$ holds, the condition $f''(R_{0}) \geq 0$ is plotted after assuming $R_{0}>0$.

\ For this model, in the zone where $G_{eff}>0$ holds, all the parameters that provide a positive scalar curvature $R_{0}>0$ also fulfill $f''(R_{0}) \geq 0$. Finally, it is possible to recover GR behavor at early times if $\alpha=1$. Hence, for this model it is also possible to obtain a constant scalar curvature solution which provides a positive contribution to Raychaudhuri equation while the model satisfies all the required conditions of viablity of Section \ref{fR_Section}. Furthermore, some of these regions are compatible with current cosmological constraints on the value of $|f'(R_{0})|$.

\bigskip

 \item {\bf Model III} $f(R)=R \left[\log(\alpha R)\right]^\beta-R$

  \smallskip

\ This model was also considered in \cite{Amendola,Black holes}. In this case, equation renders \eqref{ec_curvatura_constante}
\begin{eqnarray}
R_0=\frac{1}{\alpha}\exp(\beta)\,.
\end{eqnarray}
%

Therefore, for this model the condition guaranteeing both $R_0>0$ and a positive contribution to Raychaudhuri equation for timelike geodesics from the space-time geometry is $\alpha > 0$.
%
%

\begin{figure}
	\begin{center}
	\resizebox{8cm}{8cm}{
		\includegraphics[bb=0 0 350 350]{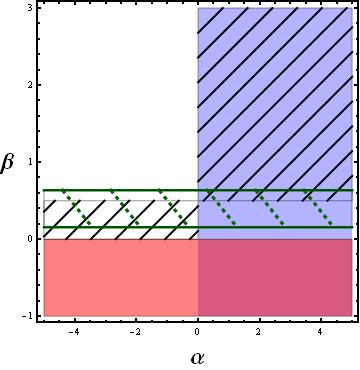}}
	\caption {\label{fig:tercermodelovacio}\footnotesize
$(\alpha,\,\beta)$ plane for Model III 
	in vacuum: As for the previous figures, $R_{0}>0$ is obtained with parameters in the blue zone. In the meshed zone $f''(R_{0})\geq0$. $G_{eff}=G/(1+f'(R_{0}))>0$ is not fulfilled in the red zone and thus the inequality \eqref{vacio} is not valid there. Let us remark that the condition $f''(R_{0})\geq0$ depends on the sign of $\beta$. As we are interested in region where $G_{eff} > 0$ holds, it is plotted the region where $f''(R_{0})\geq0$ assuming $\beta>0$. For this model, the viability condition at early times $f(R)/R\rightarrow 0$ as $R\rightarrow \infty$ does not hold for any values of the parameters. For this model, the green meshed region is compatible with the weaker upper bound on the value of $|f'(R_{0})|$, namely $|f'(R_{0})|<0.35$, but there is no region compatible with a value of $|f'(R_{0})|<0.07$.}
	\end{center}
\end{figure}

\ The same conditions considered for the previous models are plotted in Figure \ref{fig:tercermodelovacio}. For this case, there exists also a region where both conditions $R_{0}>0$ and $f''(R_{0})\geq 0$ are satisfied, namely for $\alpha > 0$ and $\beta > 1/2$. This model does not fulfill the condition $f(R)/R \rightarrow 0$ as $R \rightarrow \infty$ for any value of the parameters. However, it may be considered for late times of cosmic evolution. Moreover, there exists a region where only one of the cosmological constraint on the value of $|f'(R_{0})|$ discussed at the end of Section \ref{fR_Section} is satisfied.

\bigskip

  \item {\bf Model IV} $f(R)=-\gamma \frac{\kappa\left(\frac{R}{\gamma}\right)^n}{1+\delta \left(\frac{R}{\gamma}\right)^n}$

  \smallskip

\ This model has been proposed in \cite{Sawicki} as cosmological viable attracting
much attention in the last years. 
In order to illustrate our procedure, let us consider the particular case with $n=1$
\begin{eqnarray}
f(R)=- \frac{\alpha R}{1+\beta R}\,,
\label{ourcase}
\end{eqnarray}
where a trivial redefinition of the parameters has been performed. The constant scalar curvature of this model in vacuum becomes
\begin{eqnarray}
R_{\pm}=\frac{\alpha-1}{\beta} \pm \frac{\sqrt{\alpha (\alpha-1)}}{\beta}\,.
\label{R0Hu}
\end{eqnarray}

\begin{figure}[h!]
	\begin{center}
	\resizebox{8cm}{8cm}{
		\includegraphics[bb=0 0 350 350]{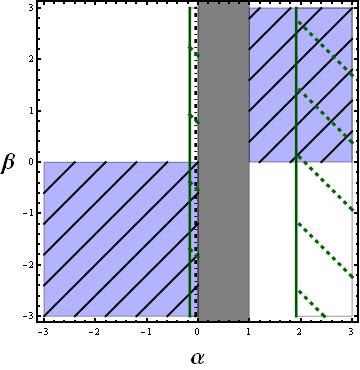}}\\
		\vspace{0.7cm}
	\resizebox{8cm}{8cm}{
		\includegraphics[bb=0 0 350 350]{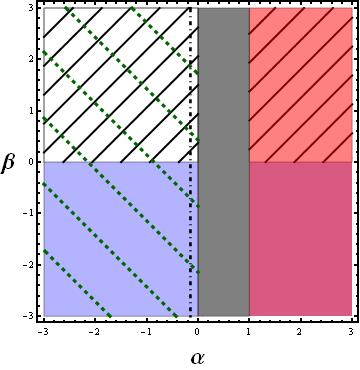}}
	\caption {\label{fig:cuartomodelovacio}\footnotesize
	$(\alpha,\,\beta)$ plane for Model IV 
	in vacuum: $R_{+}$ is considered at the upper panel whereas the lower panel considers $R_{-}$. The blue zone represents the region where $R_{+}>0$ ($R_{-}>0$). In the meshed zone the condition $f''(R_{+})\geq0$ ($f''(R_{-})\geq0$) holds. The red zone parameters can not be considered in our discussion since $G_{eff}=G/(1+f'(R_{+}))>0$ ($G_{eff}=G/(1+f'(R_{-}))>0$) does not hold there. Finally, the parameters falling in the grey zone do not fulfill $\alpha (\alpha-1) \geq 0$ which is a necessary condition in this model in order to provide a solution with constant scalar curvature. Furthermore, the viability condition $f(R)/R\rightarrow 0$ as $R \rightarrow \infty$ is always satisfied, thus this model recovers GR behavior at early times for every value of the parameters. As for the previous models, we also consider current cosmological upper bounds on the value of $|f'(R_{0})|$. The green meshed region is compatible with $|f'(R_{0})|<0.35$ and the region between the dashed black line and the grey zone is compatible with $|f'(R_{0})|<0.07$.}
	\end{center}
\end{figure}

It follows that only for $\alpha (\alpha-1) \geq 0$ constant scalar curvature solutions exist.
Therefore, imposing the constraint $R_{\pm}>0$ we get
\begin{eqnarray}
\frac{\alpha-1}{\beta} \pm \frac{\sqrt{\alpha (\alpha-1)}}{\beta} > 0\,.
\end{eqnarray}

\ Analogous plots to the previous models are shown in Figure \ref{fig:cuartomodelovacio}. For this model, two different figures are shown since there are two possible values of $R_{0}$, namely $R_{\pm}$.
For the case $R_{-}$, the regions $R_{-}>0$ and $f''(R_{-}) \geq 0$ do not overlap. On the other hand, for the case $R_{+}$, all the parameters pairs providing a positive constant scalar curvature $R_{+}>0$ also fulfill $f''(R_{+}) \geq 0$. Furthermore, for $R_{+}$ case, the condition $G_{eff}$ is always satisfied. Moreover, for the case we consider, namely \eqref{ourcase}, the condition $f(R)/R \rightarrow 0$ as $R \rightarrow \infty$ does not depend on the parameters and is always met. In the same fashion as for the previous models, it is also plotted regions where cosmological constraints on the value of $|f'(R_{0})|$ hold.

\end{itemize}

\section{Conclusions}

In this investigation we have studied the contribution of space-time geometry to the Raychaudhuri equation for timelike and null geodesics.
 In General Relativity without a cosmological constant, once the usual energy conditions are assumed it is not possible to obtain a positive contribution from the space-time geometry to the Raychaudhuri equation for timelike and null geodesics.
Nonetheless, a positive contribution from space-time geometry to the Raychaudhuri equation is obtained for many timelike directions in the present Universe \cite{ACD}.

We have proved that in $f(R)$ modified gravity theories although the same energy conditions as in General Relativity are assumed,
the fact of getting a positive contribution to the Raychaudhuri equation from space-time is allowed.

We have derived two inequalities that bound from above this contribution for congruences of both timelike and null geodesics.
In order to allow a positive contribution to the Raychaudhuri equation, these upper bounds must be positive.
The limitation with the obtained inequalities is that in general in order to extract some information, a metric solution of the modified Einstein equations must be used. Nevertheless, in the cosmological relevant case of constant scalar curvature $R_{0}$ solutions, such a knowledge is not required.
Under this assumption, 
it was obtained that $R_{ab}k^a k^b \geq 0$ where $k^a$ is a null vector. This is the condition needed for the null geodesic focusing theorem to hold.
Thus, this theorem remains valid in the context of $f(R)$ theories for space-times with constant scalar curvature, regardless the $f(R)$ model considered and the matter content (provided that the Null Energy Condition is fulfilled).
This result acquires a remarkable importance when dealing with the holographic principle in the context of $f(R)$ theories. Since if the null focusing theorem holds for this scenario in fourth order gravity theories, the light-sheets will eventually end.

Finally, in vacuum 
scenarios for space-times of constant scalar curvature $R_{0}$,
we derived constraints on the parameters of paradigmatic $f(R)$ models
guaranteeing a positive contribution to the Raychaudhuri equation. 
We conclude that for the models under consideration, there exist parameters
values that allow the desired contribution while satisfying
the imposed conditions, namely they
ensure a positive effective gravitational constant, recover of the General Relativity limit at high curvatures and guarantee the stability of the solutions.
With regard to Solar system experiments able to constrain or even discard $f(R)$ models, there is still a lack of a general formalism
applicable to all modified gravity models \cite{SolarSystem, fRcha}.
The study of gravitational waves \cite{GWaves} would also require detailed study for every model under consideration beyond the scope of this investigation.
Thus, for illustrative purposes we have only considered cosmological constraints on the value of $|f'(R_{0})|$ encountered in the literature. However, these constraints must be carefully interpreted as discussed at the end of Section \ref{fR_Section}.

For completeness, we obtain analogous conditions and arrive at the same conclusions using the alternative formulation of $f(R)$ theories as a Brans-Dicke model (see Appendix).

In this investigation, usual energy conditions were solely imposed upon the cosmological standard fluids whereas
the extra $f(R)$ terms were considered as geometrical
terms. This approach renders a more realistic analysis of the viability of $f(R)$ models
and has been studied here for the first time.
Furthermore, the sketched procedure developed  
constitutes a straightforward and systematic approach to decide whether a particular $f(R)$ model could generate
cosmological acceleration. For this purpose, we have paid particular attention to constant curvature solutions.
%
%
With the tools presented in this investigation, analysis
can be extended to more complicated cosmological scenarios
and other alternative gravity theories beyond the Concordance model \cite{ACCD}.


\begin{acknowledgments}
This work has been supported by MINECO (Spain) project numbers FIS2011-23000, FPA2011-27853-C02-01 and Consolider-Ingenio MULTIDARK CSD2009-00064. AdlCD also acknowledges financial support from Marie Curie - Beatriu de Pin\'os contract BP-B00195, Generalitat de Catalunya and ACGC, University of Cape Town.

\end{acknowledgments}


%
%

\appendix*

\section{Brans-Dicke field}

For completeness, we judged interesting to consider the implications of the inequalities \eqref{Iupper} and \eqref{IIupper} on the Brans-Dicke (BD) theories.
We can define a canonical scalar field $\phi$ in terms of the scalar curvature with the relation ($cf.$ \cite{fRcha}):
\begin{eqnarray}
f'(R)=\exp\left( \sqrt{\frac{16 \pi}{3}} \phi \right) -1\,.
\label{BDfield}
\end{eqnarray}
This field $\phi$ has associated the potential
\begin{eqnarray}
V(\phi)=\frac{R f'(R) - f(R)}{16 \pi \left( 1+f'(R) \right)^{2}}\,,
\end{eqnarray}
where $R$ depends on $\phi$ through equation \eqref{BDfield}.
Therefore, constraints \eqref{Iupper} and \eqref{IIupper} may be expressed in terms of this new field and its potential as
\begin{eqnarray}
&&{\cal M}_{\xi^a}=-R_{ab}\xi^a\xi^b \leq  \exp{\left(\sqrt{\frac{16 \pi}{3}} \phi \right)}
\nonumber\\ && \times \left[16\pi V({\phi}) - \frac{\left(2\xi^a\xi^b\nabla_a\nabla_b - \Box\right) \exp{\left(\sqrt{\frac{16 \pi}{3}} \phi \right)}}{2 \exp{\left(2 \sqrt{\frac{16 \pi}{3}} \phi \right)}}\right]\nonumber\\
\label{IupperBD}
\end{eqnarray}
and
\begin{eqnarray}
-R_{ab}k^a k^b \leq - \frac{ k^a k^b\nabla_a\nabla_b \exp{\left(\sqrt{\frac{16 \pi}{3}} \phi \right)}}{\exp{\left(\sqrt{\frac{16 \pi}{3}} \phi \right)}} \,.
\label{IIupperBD}
\end{eqnarray}
It is important to stress that we are interested in analyzing the behaviour of the curvature properties associated to the original metric of the $f(R)$ theory, dubbed as Jordan frame. Therefore, we shall not consider the conformally modified metric that characterizes the geometry in the so called Einstein frame. The reason is that free particles follow geodesics associated to the former metric and not the latter. Let us also recall that \eqref{IupperBD} and \eqref{IIupperBD} are valid provided the stress-energy tensor of the matter fields, i.e., excluding the BD field, satisfies the SEC and the NEC respectively.

In order to explore further implications of these inequalities, we can assume a constant scalar curvature space-time, as we did in the bulk of the investigation. This implies $\phi=\phi_0$ constant throughout the space-time. Thus,
\begin{eqnarray}
R_{ab}\xi^a\xi^b \geq  - 16\pi\, V({\phi_0}) \exp{\left(\sqrt{\frac{16 \pi}{3}} \phi_0 \right)}  \,,
\label{IupperBDconstant}
\end{eqnarray}
\begin{eqnarray}
R_{ab}k^a k^b \geq 0 \,;
\label{IIupperBDconstant}
\end{eqnarray}
which are the analogous equations to \eqref{Ib} and \eqref{IIb}. In particular, in Brans-Dicke formalism we have recovered that the null geodesic focusing theorem holds in space-times of constant scalar curvature and this assertion does not depend on the particular potential of the BD field. This result has important consequences when dealing with the Holographic Principle as we have already discussed in the bulk of the communication as well as in the conclusions.

For timelike geodesics, if we require a positive contribution to the Raychaudhuri equation from space-time we get the following condition to be imposed on the field potential
\begin{eqnarray}
V({\phi_0}) > 0\,,
\label{potentialcondition}
\end{eqnarray}
which is equivalent to the equation \eqref{lacondicion}. Then, in the alternative representation of $f(R)$ theories as a Brans-Dicke model, a positive contribution to the
Raychaudhuri equation for timelike geodesics is allowed if
the potential of the scalar field is positive and whatever the matter content provided its stress-energy tensor satisfies the SEC.

Furthermore, in a vacuum space-time, the scalar curvature yields
\begin{eqnarray}
R_{0}=32\pi \, V({\phi_0})\, \exp{\left(\sqrt{\frac{16 \pi}{3}} \phi_0 \right)}\,.
\end{eqnarray}
Thus, \eqref{potentialcondition} implies $R_{0}>0$ which coincides with the conclusion \eqref{Rmayor0} previously obtained for $f(R)$ theories.


\begin{thebibliography}{99}

 \bibitem{SIa}
  A.~G.~Riess {\it et al.}  [Supernova Search Team Collaboration], Astron.\ J.\  {\bf 116}, 1009 (1998);  
  S.~Perlmutter {\it et al.}  [Supernova Cosmology Project Collaboration], Astrophys.\ J.\  {\bf 517}, 565 (1999);  
  J.~L.~Tonry {\it et al.}  [Supernova Search Team Collaboration], Astrophys.\ J.\  {\bf 594}, 1 (2003). 


 \bibitem{Mod_Grav_Theories}
  A. Dobado and A. L. Maroto, Phys. Lett. B \textbf{316}, 250, (1993) [Erratum-ibid. B \textbf{321}, 435, (1994)];
  S.~Nojiri and S.~D.~Odintsov, eConf {\bf C0602061}, 06 (2006)   [Int.\ J.\ Geom.\ Meth.\ Mod.\ Phys.\  {\bf 4}, 115 (2007)];
  Phys.\ Rept.\  {\bf 505}, 59 (2011); 
  S.~Capozziello and M.~Francaviglia, Gen.\ Rel.\ Grav.\  {\bf 40}, 357 (2008); 
  T.~P.~Sotiriou and V.~Faraoni, Rev. Mod. Phys. \textbf{82} 451 (2010); 
  F.~S.~N.~Lobo, arXiv:0807.1640 [gr-qc]; 
  S. Capozziello  and  V. Faraoni, {\it Beyond Einstein Gravity}, Fundamental Theories of Physics Vol. 170, Springer Ed., Dordrecht  (2011).

 \bibitem{Lovelock}
  C.~Lanczos, Z. Phys. {\bf 73}, 147, (1932); Annals Math. {\bf 39}, 842, (1938);
  D.~Lovelock, J.\ Math.\ Phys.\  {\bf 12}, 498 (1971).

 \bibitem{GB}
  G. Cognola, E. Elizade, S. Nojiri, S. D. Odintsov and S. Zerbini, Phys. Rev. D \textbf{73} 084007 (2006); 
  S.~Nojiri, S.~D.~Odintsov, Phys. Lett. B \textbf{631} 1 (2005); 
  Phys. Rev. D \textbf{68}, 123512 (2003); 
  E.~Elizalde, R.~Myrzakulov, V.~V.~Obukhov and D.~S\'aez-G\'omez, Class. Quant. Grav. \textbf{27}  095007 (2010); 
  R.~Myrzakulov, D.~S\'aez-G\'omez and A.~Tureanu, Gen.\ Rel.\ Grav.\ \ {\bf 43} 1671 (2011); 
  A. de la Cruz-Dombriz and D. S\'aez-G\'omez, Class. Quantum Grav. {\bf 29}  245014, (2012). 

 \bibitem{ST}
  C. Brans and R. H. Dicke, Phys. Rev., {\bf 124} 925 (1961);
  C. H. Brans, Phys. Rev., {\bf 125(6)} 2194 (1962);
  J. Garc\'ia-Bellido, A. Linde and D. Linde, Phys. Rev. D {\bf 50}, 730 (1994);
  J.~A.~R.~Cembranos  {\it et al.}, JCAP {\bf 0907}, 025 (2009); 
  T.~Biswas {\it et al.}, Phys.\ Rev.\ Lett.\  {\bf 104}, 021601 (2010); 
  JHEP {\bf 1010}, 048 (2010); 
  Phys.\ Rev.\  D {\bf 82}, 085028 (2010). 

 \bibitem{VT}
  L.~H.~Ford, Phys.\ Rev.\ D {\bf 40} (1989) 967.
  J.~Beltr\'an Jim\'enez and A.~L.~Maroto, Phys.\ Rev.\ D {\bf 78} (2008) 063005; JCAP {\bf 0903} (2009) 016;
  Phys.\ Rev.\ D {\bf 80} (2009) 063512;
  T.~Koivisto and D.~F.~Mota, JCAP {\bf 0808}, 021 (2008); 
  J.~A.~R.~Cembranos  {\it et al.}, Phys.\ Rev.\ D {\bf 86}, 021301 (2012); 
  arXiv:1212.3201 [astro-ph.CO]. 

 \bibitem{XD}
  J.~Alcaraz {\it et al.}, Phys. Rev. D {\bf 67}, 075010 (2003); 
  P. Achard {\it et al.}, Phys. Lett. B {\bf 597}, 145 (2004); 
  J.~A.~R.~Cembranos, A.~Dobado and A.~L.~Maroto,  Phys.\ Rev.\ Lett.\  {\bf 90}, 241301 (2003); 
  Phys.\ Rev.\ D {\bf 68}, 103505 (2003); 
  AIP Conf.Proc. {\bf 670}, 235 (2003); 
  Int. J. Mod. Phys. D {\bf 13}, 2275 (2004); 
  Phys. Rev. D {\bf 70}, 096001 (2004); 
  Phys.\ Rev.\ D {\bf 73}, 035008 (2006); 
  Phys.\ Rev.\ D {\bf 73}, 057303 (2006); 
  J.\ Phys.\ A  {\bf 40}, 6631 (2007). 

\bibitem{sugra}
  D.~Z.~Freedman, P.~van Nieuwenhuizen and S.~Ferrara, Phys.\ Rev.\ D {\bf 13}, 3214 (1976); 
  S.~Deser and B.~Zumino, Phys.\ Lett.\ B {\bf 62}, 335 (1976);  
  E.~Cremmer, B.~Julia and J.~Scherk, Phys.\ Lett.\ B {\bf 76}, 409 (1978); 
  L.~J.~Hall, J.~D.~Lykken and S.~Weinberg, Phys.\ Rev.\ D {\bf 27}, 2359 (1983); 
  N.~Ohta, Prog.\ Theor.\ Phys.\  {\bf 70}, 542 (1983); 
  L.~Alvarez-Gaume, J.~Polchinski and M.~B.~Wise,  Nucl.\ Phys.\ B {\bf 221}, 495 (1983); 
  H.~P.~Nilles, Phys.\ Rept.\  {\bf 110}, 1 (1984). 
  J.~A.~R.~Cembranos, J.~L.~Feng, A.~Rajaraman and F.~Takayama, Phys.\ Rev.\ Lett.\  {\bf 95}, 181301 (2005); 
  AIP Conf.\ Proc.\  {\bf 903}, 591 (2007). 
  J.~A.~R.~Cembranos, J.~L.~Feng and L.~E.~Strigari, Phys.\ Rev.\ Lett.\  {\bf 99}, 191301 (2007); 
  Phys.\ Rev.\  D {\bf 75}, 036004 (2007); 
  M.~R.~Garousi, arXiv:1210.4379 [hep-th].

 \bibitem{disformal}
  G.~W.~Horndeski, Int.\ J.\ Theor.\ Phys.\  {\bf 10}, 363 (1974);  
  J.~D.~Bekenstein, Phys.\ Rev.\ D {\bf 48}, 3641 (1993); 
  J.~A.~R.~Cembranos {\it et al.}, Phys. Rev. D {\bf 65} 026005 (2002); 
  JCAP {\bf 0810}, 039 (2008); 
  Phys.\ Rev.\  D {\bf 83}, 083507 (2011); 
  Phys.\ Rev.\ D {\bf 84}, 083522 (2011); 
  Phys.\ Rev.\ D {\bf 85}, 043505 (2012); 
  arXiv:1204.0655 [hep-ph]. 
  J.~A.~R.~Cembranos and L.~E.~Strigari, Phys.\ Rev.\  D {\bf 77}, 123519 (2008); 
  M.~Zumalacarregui, T.~S.~Koivisto, D.~F.~Mota and P.~Ruiz-Lapuente, JCAP {\bf 1005}, 038 (2010); 
  T.~S.~Koivisto, D.~F.~Mota and M.~Zumalacarregui, arXiv:1205.3167 [astro-ph.CO]; 
  arXiv:1210.8016 [astro-ph.CO]. 

 \bibitem{LV}
  V. A. Kostelecky and S. Samuel, Phys. Rev. D {\bf 39}, 683 (1989).
  D. Colladay and V. A. Kostelecky, Phys. Rev. D {\bf 55}, 6760 (1997);
  J. R. Ellis, N. E. Mavromatos and D. V. Nanopoulos,  Phys. Rev. D {\bf 61}, 027503 (1999);
  J. Alfaro, H. A. Morales-Tecotl and L. F. Urrutia, Phys. Rev. Lett. {\bf 84}, 2318 (2000);
  G. Amelino-Camelia, Nature {\bf 418}, 34 (2002);
  J. Magueijo and L. Smolin, Phys. Rev. Lett. {\bf 88}, 190403 (2002);
  J.~A.~R.~Cembranos, A.~Rajaraman and F.~Takayama, hep-ph/0512020; 
  Europhys.\ Lett.\  {\bf 82}, 21001 (2008); 
  S. Ghosh and P. Pal, Phys. Rev. D {\bf 75}, 105021 (2007).

 \bibitem{reviewsf(R)}
  T.~P.~Sotiriou, J.\ Phys.\ Conf.\ Ser.\  {\bf 189}, 012039 (2009); 
  S.~Capozziello and M.~De Laurentis, Phys.\ Rept.\  {\bf 509}, 167 (2011); 
  S.~'i.~Nojiri and S.~D.~Odintsov, Phys.\ Rept.\  {\bf 505}, 59 (2011); 
  T.~P.~Sotiriou and V.~Faraoni, Rev.\ Mod.\ Phys.\  {\bf 82}, 451 (2010);  
  A.~de la Cruz-Dombriz and D.~S\'aez-G\'omez, Entropy {\bf 14}, 1717 (2012).

 \bibitem{tests}
  T.~P.~Sotiriou, Gen.\ Rel.\ Grav.\  {\bf 38}, 1407 (2006); 
  S.~'i.~Nojiri and S.~D.~Odintsov, Phys.\ Rept.\  {\bf 505}, 59 (2011); 
  B.~Jain and J.~Khoury, Annals Phys.\  {\bf 325}, 1479 (2010); 
  A. de la Cruz-Dombriz, A. Dobado and A. L. Maroto, Phys.\ Rev. D {\bf 77}, 123515 (2008);
  Phys.\ Rev.\ Lett. {\bf 103} 179001 (2009);
  M. Abdelwahab, A. Abebe, A. de la Cruz Dombriz and P. K. S. Dunsby, Class.\ Quant.\ Grav. {\bf 29}  135011 (2012);
  J. A. R. Cembranos {\it et al.}, JCAP {\bf 1204}, 021 (2012); 
  AIP Conf. Proc. {\bf 1458}  491 (2011);
  H. Bourhrous, A. de la Cruz Dombriz and P. K. S. Dunsby, AIP Conf. Proc. {\bf 1458} 343, (2011).
  A.~Abebe, A.~de la Cruz-Dombriz and P.~K.~S.~Dunsby, arXiv:1304.3462 [astro-ph.CO];
  F.~G.~Alvarenga, A.~de la Cruz-Dombriz, M.~J.~S.~Houndjo, M.~E.~Rodrigues and D.~S\'aez-G\'omez,  arXiv:1302.1866 [gr-qc].





 \bibitem{HE}
  S.~W.~Hawking and G.~F.~R.~Ellis, \textit{The large scale structure of space-time},
  (Cambridge University Press, Cambridge, U.K., 1973).

 \bibitem{Wald}
  R.~M.~Wald, \textit{General Relativity}, (University of Chicago Press, Chicago, U.S.A., 1984).

 \bibitem{Raychaudhuri}
  A.~Raychaudhuri, Phys.\ Rev.\  {\bf 98}, 1123 (1955).

 \bibitem{Sachs}
  R.~K.~Sachs, Proc.\ Roy.\ Soc.\ Lond.\ A {\bf 264}, 309 (1961).

 \bibitem{Ehlers}
  J.~Ehlers, Gen. Rel. Grav. {\bf 25}, 1225 (1993), English translation of original 
  article by P. Jordan, J. Ehlers, W. Kundt, R. K. Sachs, Proceedings of the Mathematical Natural Sciences Section of the Mainz
  Academy of Sciences and Literature, Nr. {\bf 11}, 792 (1961).

 \bibitem{SayanKarandDadhich}
  G.~F.~R.~Ellis, Pramana {\bf 69}, 15 (2007);
  S.~Kar and S.~SenGupta, Pramana {\bf 69}, 49 (2007); 
  N.~Dadhich, gr-qc/0511123.

 \bibitem{ACD}
  F.~D.~Albareti, J.~A.~R.~Cembranos and A.~de la Cruz-Dombriz, JCAP {\bf 1212}, 020 (2012). 

 \bibitem{Eisenhart}
  L.~P.~Eisenhart, \textit{Riemannian Geometry}, (Princeton University Press, Princeton, New Jersey, U.S.A., 1926).

 \bibitem{Santos}
  J.~Santos, J.~S.~Alcaniz, M.~J.~Reboucas and F.~C.~Carvalho, Phys.\ Rev.\ D {\bf 76}, 083513 (2007); 
  J.~Santos, M.~J.~Reboucas and J.~S.~Alcaniz, Int.\ J.\ Mod.\ Phys.\ D {\bf 19}, 1315 (2010).  

 \bibitem{Atazadeh}
  K.~Atazadeh, A.~Khaleghi, H.~R.~Sepangi and Y.~Tavakoli, Int.\ J.\ Mod.\ Phys.\ D {\bf 18}, 1101 (2009). 

 \bibitem{Bertolami}
  O.~Bertolami and M.~C.~Sequeira, Phys.\ Rev.\ D {\bf 79}, 104010 (2009). 

 \bibitem{Garcia}
  N.~M.~Garcia, T.~Harko, F.~S.~N.~Lobo and J.~P.~Mimoso, Phys.\ Rev.\ D {\bf 83}, 104032 (2011); N.~Montelongo Garcia, F.~S.~N.~Lobo, J.~P.~Mimoso and T.~Harko,
  J.\ Phys.\ Conf.\ Ser.\  {\bf 314}, 012056 (2011).
  
 \bibitem{Banijamali}
  A.~Banijamali, B.~Fazlpour and M.~R.~Setare, Astrophys.\ Space Sci.\  {\bf 338}, 327 (2012). 

 \bibitem{Zhao}
  Y.~-Y.~Zhao, Y.~-B.~Wu, J.~Lu, Z.~Zhang, W.~-L.~Han and L.~-L.~Ling, Eur.\ Phys.\ J.\ C {\bf 72}, 1924 (2012). 

 \bibitem{Dombriz-Dunsby-Odintsov}
  A.~de la Cruz-Dombriz and A.~Dobado, Phys.\ Rev.\ D {\bf 74}, 087501 (2006); 
  P.~K.~S.~Dunsby, E.~Elizalde, R.~Goswami, S.~Odintsov and D.~S.~Gomez, Phys.\ Rev.\ D {\bf 82}, 023519 (2010); 
  S.~'i.~Nojiri and S.~D.~Odintsov, Phys.\ Rev.\ D {\bf 74}, 086005 (2006). 

 \bibitem{Dolgov-Kawasaki}
  A.~D.~Dolgov and M.~Kawasaki, Phys.\ Lett.\ B {\bf 573}, 1 (2003). 

 \bibitem{Clifton}
  T.~Clifton, P.~G.~Ferreira, A.~Padilla and C.~Skordis, Phys.\ Rept.\  {\bf 513}, 1 (2012). 

 \bibitem{Cauchy}
  S.~Capozziello and S.~Vignolo, Int.\ J.\ Geom.\ Meth.\ Mod.\ Phys.\  {\bf 9}, 1250006 (2012). 

 \bibitem{Pogosian}
  L.~Pogosian and A.~Silvestri, Phys.\ Rev.\ D {\bf 77}, 023503 (2008) [Erratum-ibid.\ D {\bf 81}, 049901 (2010)]. 

 \bibitem{fRcha}
  J.~A.~R.~Cembranos, Phys.\ Rev.\  D {\bf 73}, 064029 (2006). 

 \bibitem{Referee}
  B.~Jain, V.~Vikram and J.~Sakstein,
  arXiv:1204.6044 [astro-ph.CO];
  L.~Lombriser, A.~Slosar, U.~Seljak and W.~Hu,
  Phys.\ Rev.\ D {\bf 85}, 124038 (2012);
  L.~Lombriser, F.~Schmidt, T.~Baldauf, R.~Mandelbaum, U.~Seljak and R.~E.~Smith,
  Phys.\ Rev.\ D {\bf 85}, 102001 (2012); 
  Y.~-S.~Song, H.~Peiris and W.~Hu,
  Phys.\ Rev.\ D {\bf 76}, 063517 (2007).

 \bibitem{Hu}
  Y.~-S.~Song, W.~Hu and I.~Sawicki,
  Phys.\ Rev.\ D {\bf 75}, 044004 (2007).

 \bibitem{Bousso}
  R.~Bousso, Rev.\ Mod.\ Phys.\  {\bf 74}, 825 (2002). 


 \bibitem{Carroll}
  S.~M.~Carroll, V.~Duvvuri, M.~Trodden and M.~S.~Turner, Phys.\ Rev.\ D {\bf 70}, 043528 (2004). 

\bibitem{485} V. Faraoni, 
Phys. Rev. D, 74 10 104017 (2006).

 \bibitem{Starobinsky}
  A.~A.~Starobinsky, Phys.\ Lett.\ B {\bf 91}, 99 (1980).

 \bibitem{Cembranos}  J.~A.~R.~Cembranos, Phys.\ Rev.\ Lett.\  {\bf 102}, 141301 (2009). 

 \bibitem{26} C. P. L. Berry and J. R. Gair, Phys. Rev. D {\bf 83} 104022 (2011).

 \bibitem{varia_Rn} 
A. L. Erickcek, T. L. Smith, and M. Kamionkowski, 
Phys. Rev. D, {\bf 74} 12 121501 (2006); 
M. Fairbairn and S. Rydbeck. 
Astropart. Phys., {\bf 12} 5 (2007); 
I. Sawicki and W. Hu, 
Phys. Rev. D {\bf 75}, 12 127502 (2007).



 \bibitem{Amendola}
  L.~Amendola, R.~Gannouji, D.~Polarski and S.~Tsujikawa, Phys.\ Rev.\ D {\bf 75}, 083504 (2007). 

 \bibitem{Black holes}
  A.~de la Cruz-Dombriz, A.~Dobado and A.~L.~Maroto, Phys.\ Rev.\ D {\bf 80}, 124011 (2009) [Erratum-ibid.\ D {\bf 83}, 029903 (2011)]; 
  J.~A.~R.~Cembranos, A.~de la Cruz-Dombriz and P.~J.~Romero,   arXiv:1109.4519 [gr-qc]; 
  AIP Conf.\ Proc. {\bf 1458} 439 (2011).

 \bibitem{Sawicki}
  W.~Hu and I.~Sawicki, Phys.\ Rev.\ D {\bf 76}, 064004 (2007). 

 \bibitem{SolarSystem}
    V.~Faraoni,  
  Phys.\ Rev.\ D {\bf 74}, 023529 (2006); 
  %
   I.~Navarro and K.~Van Acoleyen,   
  JCAP {\bf 0603}, 008 (2006); 
  %
  G.~Calcagni, B.~de Carlos and A.~De Felice,   
  Nucl.\ Phys.\ B {\bf 752}, 404 (2006); 
%
  G.~J.~Olmo, 
  Phys.\ Rev.\ D {\bf 72}, 083505 (2005); 
  Phys.\ Rev.\ Lett.\  {\bf 95}, 261102 (2005); 
 %
  S.~Capozziello, V.~F.~Cardone and A.~Troisi,  
  Mon.\ Not.\ Roy.\ Astron.\ Soc.\  {\bf 375}, 1423 (2007). 


 \bibitem{GWaves}   C.~P.~L.~Berry and J.~R.~Gair,
  Phys.\ Rev.\ D {\bf 83}, 104022 (2011)   [Erratum-ibid.\ D {\bf 85}, 089906 (2012)].

 \bibitem{ACCD}
  F.~D.~Albareti, J.~A.~R.~Cembranos, A.~de la Cruz-Dombriz and A.~Dobado, in progress.


\end{thebibliography}
\end{document}